\documentstyle[12pt]{article}

\def\mxth{\mathsurround=0pt }
\def\xversim#1#2{\lower2.pt\vbox{\baselineskip0pt \lineskip-.2pt
    \ialign{$\mxth#1\hfil##\hfil$\crcr#2\crcr\sim\crcr}}}

\def\gtsim{\mathrel{\mathpalette\xversim >}}

\begin{document}
\begin{flushright}UPR-0510T\end{flushright}
\vspace*{.5in}
\begin{center}
{\bf GRACEFUL EXIT IN EXTENDED INFLATION AND IMPLICATIONS FOR DENSITY
PERTURBATIONS}\\
\vspace{.4in}
 Robert Crittenden and Paul J. Steinhardt \\
{\it Department of Physics \\
University of Pennsylvania \\
Philadelphia, PA  19104}
\end{center}
\pagestyle{empty}

\vspace{.1in}
\noindent
{\bf ABSTRACT:}  Two qualitatively different modes of ending superluminal
expansion are possible in extended inflation.  One mode, different
from the one envoked in most extended models to date,
easily avoids making big bubbles
that  distort the cosmic microwave background radiation (CMBR).  In this mode,
the   spectrum of density fluctuations
  is found to be scale-free, $P(k) \propto k^n$, where $n$ might lie anywhere
between 0.5 and 1.0  (whereas, previously, it appeared that
the range $1.0> n \gtsim 0.84$ was
disallowed).

\newpage
\pagestyle{plain}
\setcounter{page}{1}

Extended inflation$^{1,2}$ was introduced to resolve the fine-tuning problem
that
plagued previous inflationary models.  As in ``old inflation,"$^3$ the scale
factor $a(t)$ begins to expand superluminally ($a(t)>t$) when
the universe
supercools  into a
metastable false vacuum
phase during a strongly first order phase transition.
Unlike old inflation, though, graceful exit can be  achieved by
bubble nucleation,
quantum tunneling through the energy barrier.  The added element is
modifications to Einstein gravity that slow the expansion from
exponential ($a(t) = {\rm exp}(Ht)$ and constant Hubble parameter $H$) to
polynomial ($a(t)>t$, but $H$ steadily decreasing).  For
slowed expansion, the dimensionless parameter $\epsilon \equiv \Gamma/H^4$ ---
which measures the competition between the nucleation rate
$\Gamma$ and the expansion rate --- steadily increases.$^1$ Initially,
$\epsilon \ll 1$ and inflation dominates nucleation.  In old inflation,
$\epsilon$ is
constant  so inflation never stops.  In extended
inflation,  $\epsilon$
steadily increases until $\epsilon >1$, at which point nucleation dominates
expansion and the transition to the true vacuum is completed.

The purpose of this paper is to show that extended inflation offers two
qualitatively distinct modes of ``graceful exit."   In the mode
considered by most authors, inflation stops when bubbles of true vacuum
consume the false vacuum. This mode can create problems of its own,
including unacceptable inhomogeneities created by big bubbles.$^{4,5}$  Our
point
here is to emphasize a  second mode in which inflation ends
while the universe is still trapped in the false phase.  We argue that this
mode automatically avoids the big bubble problem. We also
examine the implications
for the
the density perturbation spectrum.
 In a spectrum characterized
by a power-law, $P(k) \propto k^n$,
scale invariance (a Harrison-Zel'dovich  spectrum) corresponds to $n=1$.
Previously, avoiding the big bubble problem seemed to require $n < 0.84$, and,
hence, a spectrum tilted away from strict scale-invariance.  However,
with the novel graceful exit mechanism, the range opens up to $1.0 \ge n
\gtsim 0.5$, including $n$ arbitrarily close to  unity.

  For the purposes of illustration, we will consider modified gravity theories
in which a scalar field $\phi$ is non-minimally coupled
to the scalar curvature, ${\cal R}$:
 \begin{equation}
{\cal L}  =  - f(\phi) {\cal R} - \frac{1}{2}(\partial_{\mu} \phi)^{2} -
 \frac{1}{2}(\partial_{\mu} \sigma)^{2} -V(\sigma) + {\cal L}_{other}.
\end{equation}
 The field $\sigma$
is an  order parameter (e.g., a Higgs field) for
a strongly first order transition; during supercooling, $\sigma$ is
trapped in a false vacuum phase with energy
density is $V(\sigma) = M_{F}^{4}$.
 ${\cal L}_{other}$ includes all other
matter fields and any potential for $\phi$.
It is useful to recast the action  in terms of a new scalar field,
 $\Phi \equiv f(\phi)$ and a function
$
\omega(\phi) \equiv \frac{f(\phi)}{2 [f'(\phi)]^2}
$:
\begin{equation}
{\cal L} = - \Phi {\cal R} - \omega(\Phi) \frac{(\partial_{\mu}
\Phi)^{2}}{\Phi}
-
 \frac{1}{2}(\partial_{\mu} \sigma)^{2} -V(\sigma) + {\cal L}_{other}.
 \end{equation}
 The original Brans-Dicke model corresponds to
 $f(\phi)= \xi \phi^2$,  a special case
 in which $\omega(\Phi) = 1/8 \xi$ is constant.$^{1,6}$  The  generic case is
where $\omega$ is $\Phi$-dependent,$^2$ a generalization which leads to two
distinct modes of ending extended inflation:

\noindent
{\it Mode 1: Inflation ends by true vacuum bubbles consuming the
false vacuum.} This mode, considered in the first extended inflation
papers,$^{1,2,4,5,7}$
 is exemplified by
 the original
Brans-Dicke theory with constant $\omega$.  While the universe is trapped
in the false vacuum,
 the scale factor increases as a power-law,$^1$ $a(t) \propto t^{\omega +
\frac{1}{2}}$, and the Hubble parameter,
$H = \dot{a}/a = (\omega+ \frac{1}{2})/t$,
decreases uniformly.  Hence, $\epsilon =\Gamma/H^4$ increases steadily, as
desired.

Density perturbations are generated during inflation from two sources.  Quantum
fluctuations in $\phi$ result in a nearly scale-invariant spectrum:$^{7,8}$
\begin{equation}
\frac{\delta \rho}{\rho}|_{H} \,(\lambda) \approx \frac{M_{F}^2}{m_P^2}
\;g(\omega)
\; \lambda^{4/(2\omega-1)},
\end{equation}
where $\frac{\delta \rho}{\rho}|_{H}(\lambda)$ is the perturbation amplitude as
length $\lambda$ re-enters the horizon after inflation;
$m_P^2 \equiv f(\phi)$ is the effective Planck mass at the end of
superluminal expansion; and $g(\omega)$ is an $\omega$-dependent
correction$^8$ that is close to unity except for $\omega \rightarrow 1$.
Note that a $\frac{\delta \rho}{\rho}|_{H}$ spectrum which scales as
$\lambda^m$
converts to a power spectrum $P(k) \propto k^{1-2m}$.
For $M_F/m_P \approx 10^{-(3-5)}$, but with no  fine-tuning of
couplings, $\frac{\delta \rho}{\rho}|_{H} (\lambda)$ is consistent with recent
COBE observations.
A
second  source of inhomogeneities comes from big bubbles.$^{4,5}$  If inflation
continues until bubbles consume all remaining false vacuum,
the distribution of bubble sizes at the end of inflation is $
F(x>x_0) \approx \frac{1}{(1 + H_e x_0)^{4/\omega}}$,
 where $F(x>x_0)$ is the fractional volume occupied by bubbles of
 proper radius greater than $x_0$ and $H_{e}$ is the Hubble parameter
 at the end of inflation.$^5$  To avoid  distortion of the CMBR, we require
$F(x>x_0)$ to be less than $10^{-4}$ for
 bubbles of supercluster scale or larger, $H_{e} x_0 \ge 10^{25}$;  this
 implies $\omega < 25$.

 The serious criticism of this mode is
  the disappointingly limited
 range of workable models.
 Although the phase transition can be completed
 for any $\omega < \infty$, only $ \omega <25$  leads to
  acceptable inhomogeneities.   (Observational constraints on Brans-Dicke
 theory conflict with $\omega <25$ if $\phi$ is massless, but are
 avoided by assuming a finite
 mass $\gtsim 10^{-12}$~GeV.$^{9}$)

\noindent
{\it Mode 2:  Inflation ends while the universe remains trapped in the false
phase.}  Generically,
$\omega$ is $\phi$-dependent and time-dependent, enabling
a new mechanism for terminating superluminal expansion.
For
a very wide range of  polynomial or
exponential $f(\phi)$, $\omega$ decreases sharply with increasing $\phi$.
See Figure 1.
  When $\phi$ is small, $\omega >>1$ and
the universe expands superluminally.  Through
 the non-minimally coupling term, the false
vacuum energy density drives $\phi$ to steadily increase.
As $\phi$ grows, $\omega$ decreases and $a(t)$  accelerates
less rapidly.
 The boundary, $\omega = 1/2$, is critically important.
Recall that the scale factor grows as $a(t) \approx t^ {\omega + \frac{1}{2}}$,
for constant $\omega$.$^1$  The expansion becomes subluminal ---
non-inflationary --- once $\omega$ drops below 1/2, even though
the universe remains
trapped in the false vacuum.  Even after this point,
$\phi$ continues to grow, the expansion rate
continues to decrease, and the bubble nucleation parameter, $\epsilon$,
continues to increase.  Some time after inflation has
stopped,  $\epsilon$ finally increases above unity and bubble nucleation ends
the transition to the false phase.  As in Mode 1, several mechanisms$^5$ are
possible
which can later pin $\phi$ at
a fixed value (e.g., $\phi$ gets a finite mass).

Although this possibility was briefly mentioned
in  Ref. 2, we have recently come to see this as a truly
 distinct mode of graceful exit  that evades some of
the criticisms of earlier models.   First, our recent
numerical calculations
suggest that viable models exist for a wide range of functional forms and
parameters, significantly wider than Mode 1.
(See Figs. 2 and 3 for some examples.)
Second, the big bubble problem is easily avoided because few bubbles
are nucleated ($\epsilon \ll 1$) while the universe is inflating.
(In Mode 1,the troublesome CMBR distortion comes from  inflation
of the many bubbles nucleated as $\epsilon$ gets close to unity.$^5$)
 Almost the entire false vacuum is
converted to true vacuum by bubbles nucleated after the expansion has
become subluminal.  These uninflated bubbles   are
infinitessimally smaller than the horizon
at decoupling, much too small to distort the CMBR.
There remains one criticism: that quartic interactions in any
potential for $\phi$ must be
suppressed by an amount comparable to the fine-tuned  interactions
in new or chaotic inflation.$^{10}$  The  fine-tuned
interactions in new or chaotic
inflation are essential elements that actually drive the inflationary
expansion;
they must be fine-tuned  into a tiny range that must not include zero.
Here, we would emphasize that
the quartic interactions in $\phi$ play no role in inflation, and,
hence, it is acceptable if symmetries prohibit them altogether.  (Such
symmetries exist in superstring models, for example.) Hence, it is difficult
to judge how serious this last criticism is to be taken.

We turn, then,
to  the density perturbation spectrum generated by this novel
mechanism for terminating inflation.  According to the recent COBE
results,$^{11}$
the  CMBR anisotropy can be
fit to a scale-free
power spectrum, $P(k) \propto k^n$, where $n= 1.1 \pm 0.6$ ($1 \sigma$
limit).
Exponent
$n=1$ corresponds to an exactly
scale-invariant, Harrison-Zeldovich spectrum.   New and chaotic
inflation
result in a nearly
 scale-invariant spectrum with logarithmic deviations
 ($\propto
{\rm log} \; \lambda$).  We have computed the exponent $n$ by fitting the
density
perturbation spectrum to a scale-free form
over astrophysical scales, $1-10^4$~Mpc.    For new or chaotic
inflation, our fit gives $n \approx 0.95.$
For constant $\omega$, the density perturbation spectrum  in
Eq.~(3) converts to a power spectrum with
\begin{equation}\label{n}
n = \frac{2 \omega -9}{2 \omega-1}.
\end{equation}
For large $\omega$ (approaching the limit of
Einstein gravity), the spectrum is  near $n=1$.
As $\omega$ decreases, the
spectrum shifts to smaller values of $n$,
tipping the spectrum towards more power on large scales.

For Mode~1, we argued that $\omega$ must be less than 25 to avoid the big
bubble problem.
According to Eq.~(\ref{n}),  this requires $n<0.84$, a significant tilt away
from
scale-invariance.

For Mode~2, a greater range of  $n$ is possible.
In general, the amplitude $\frac{\delta \rho}{\rho}|_{H}(\lambda)$
is $\approx H^2/\dot{\phi}$
evaluated as scale $\lambda$ is stretched beyond the horizon during inflation.
In Mode~2, $\omega$, which
controls the rate of inflation, varies rapidly during inflation, especially
within the final e-foldings. In this case, there is no simple analytic
expression for the amplitude; rather, $\phi$, $\omega$, $H$, and
$\frac{\delta \rho}{\rho}|_{H}$ have to be tracked by numerically solving the
equations of
motion.
Our results are summarized in Figs. 2-4.
To compare with COBE,$^{11}$ we  fit  the upper curves ($1-10^4$~Mpc)
to a scale-free
spectrum and determine $n$.  Fluctuations on these scales were
generated 50-60 e-foldings before the end of inflation and depend on
$\omega$ during that period.$^{5,7,8}$

We find three classes of behavior, depending on the functional
form for the non-minimal coupling, $f(\phi)$:
(1)~If $f(\phi)$ varies exponentially
with $\phi$ (see Fig.~2),  then $\omega$ remains very large until the
last 10 e-foldings of inflation.  Consequently, on scales $1-10^4$~Mpc,
the spectrum is flat, $n \rightarrow 1$, virtually indistinguishable from
new inflation. This regime was strictly disallowed in the original
(Mode~1) graceful exit mechanism for extended inflation.
  (2)~If $f(\phi)$ varies more slowly with $\phi$ (e.g.,
a polynomial), $\omega$ changes more gradually.  Hence, $\omega$ is
closer to 1/2  during the last 50-60 e-foldings of
inflation compared to (1), leading to a spectrum tipped towards $n<1$ on scales
$1-10^4$~Mpc.
However, $\omega$ cannot be too small and still satisfy all other
inflationary constraints.$^5$
We find that   $1 \gtsim n \gtsim 0.5$ is spanned by a plausible range
of parameters and  polynomial forms.
 (Yet smaller values of $n$ are possible, in principle, but only
if $f(\phi)$ is increasingly fine-tuned.)
(3)~Wild
variations  can be ``designed'' into the spectrum by special choices of
$f(\phi)$.
For example, $\omega(\phi) \equiv f/2[f']^2$ blows up at any near-inflection
point of $f(\phi)$. Consequently,
sharp ``dips'' in the perturbation spectrum
(see Fig.~4) are created as $\phi$ passes the near-inflection point.
The dips are similar to features created in ``designer
inflation'' models$^{12,13}$ in new or chaotic inflation.
By choosing parameters carefully, the dip can be made to lie within
$1-10^4$~Mpc
scales (Fig.~4a), which is probably ruled out by COBE.   If the dip is placed
just beyond the
horizon, the spectrum on $1-10^4$~Mpc scales appears
to be scale-free (no dips), tipped towards $n>1$. Values
$1.3  \gtsim n \ge 1.0$ can be obtained.
See Fig.~ 4b.  What should be emphasized, though,
 is that
cases (1) and (2) occur for a robust set of functional forms and parameters,
whereas case (3) only occurs for a very special choice of $f(\phi)$.

Our conclusion is that  extended
inflation, based on the mode in which expansion becomes
subluminal before escape from  the false vacuum, is  viable and robust.
  The  power spectrum for extended inflation is scale-free with a
wider range of possible power indices than previously thought possible:
$1.0 \ge n \gtsim 0.5$ for  plausible models, tilted towards
 more power on large scales ($n=1$ is exact Harrison-Zel'dovich).
Recently, it has been shown that extended inflation
models with tilts $0.84\gtsim n \gtsim 0.5$ have the
additional property that gravitational waves, rather than
density perturbations, are the dominant contribution to cosmic
microwave background fluctuations on large angular scales.$^{14}$
  The range
 $n>1$ or strong deviations from a scale-free spectrum require unnatural
   functional forms and fine-tuning.  The predicted range is within the
  present $1\sigma$-limit for COBE.$^{11}$  However, future
  experiments, including analysis of years two or more from COBE, should
  improve the measurements and provide useful new
  constraints on inflation models.

 \newpage
\begin{center}
{\bf FIGURE CAPTIONS}
\end{center}
\pagestyle{empty}

\noindent
1. Typical behavior of
$\omega(\phi)= f/2[f']^2$ for polynomial or exponential
$f(\phi)$.  In Mode 2,
extended inflation begins when $\omega \gg 1$
and ends when $\omega$ falls below 1/2. Bubble nucleation is negligible
until $\omega \ll 1/2$, too late for inflation to make many big bubbles.

\noindent
2.
$\delta \rho/\rho|_H$ as wavelength $\lambda$ enters the horizon
shown for
all observable scales (lower), with  blow-up showing large
scales  only (upper).  Scale-invariance or, equivalently,
a power spectrum with $n=1$, corresponds to a horizontal line.
 For  exponential
$f(\phi)$, [here, $f(\phi)= M^2 {\rm exp} (.048 \phi/M)$, where
$\phi \le M$ initially and $M\gtsim M_F$], the spectrum
on large scales can be fit to $n\approx 1$  [here, $n=0.96$].

\noindent
3.  Same as Fig.~2 except for polynomial $f(\phi)$ [here, $f(\phi)=
A (\phi^2 + B \phi^4/M^2)$, where $A=0.0061$ and $B=0.0021$].   The best-fit to
upper
curve is $n=0.8$.
Values $1.0 \ge n \gtsim 0.5$ can be obtained by varying $A$ and $B$, but
$B/A$ must be increasingly tuned for yet smaller $n$.

\noindent
4.  Same as Fig.~2 except $f(\phi)= A (3.125 \, \delta \, \phi^2 -
1.66  \phi^3/M^2 + 0.25 \phi^4)$
is designed to have an inflection point  as $\delta \rightarrow 1$.
Depending on parameters, a sharp dip may be created
 on astrophysical scales (lower and upper left; $A=0.2$ and
 $\delta=1.0001$);
or,  it can be placed just beyond the horizon
  $\lambda \approx 10^4$~Mpc (upper right; $A= 0.059$ and
  $\delta=1.001$), which appears  scale-free within the
  horizon with $n\approx 1.3$.

\end{document}